Draft Accepted by ApJ

# The Decrease of Specific Angular Momentum and the Hot Toroid Formation: The Massive Clump G10.6-0.4

Hauyu Baobab Liu[1,2,3]

*Harvard-Smithsonian Center for Astrophysics, 60 Garden Street, Cambridge, MA 02138*

hlu@cfa.havard.edu

Paul T. P. Ho[1,3]

*Academia Sinica Institute of Astronomy and Astrophysics,*
*P.O. Box 23-141, Taipei, Taiwan*

pho@asiaa.sinica.edu.tw

Qizhou Zhang[1]

Eric Keto[1]

*Harvard-Smithsonian Center for Astrophysics, 60 Garden Street, Cambridge, MA 02138*

qzhang@cfa.harvard.edu

keto@cfa.harvard.edu

Jingwen Wu[5]

*Jet Propulsion Laboratory, 4800 Oak Grove Dr., Pasadena, CA, 91101*

Jingwen.Wu@jpl.nasa.gov

Huabai Li[4]

*Max Planck Institute for Astronomy, Königstuhl 17, D-69117 Heidelberg, Germany*

li@mpia.de

[1]Harvard-Smithsonian Center for Astrophysics

[2]Department of Physics, National Taiwan University

[3]Academia Sinica Institute of Astronomy and Astrophysics

[4]Max Planck Institute for Astronomy

[5]Jet Propulsion Laboratory

## ABSTRACT

This is the first paper of our series of high resolution (1″) studies of the massive star forming region G10.6–0.4. We present the emission line observations of the hot core type tracers ($O^{13}CS$, OCS, $SO_2$) with $\sim 0''.5$ resolution. By comparing the results to the high–resolution $NH_3$ absorption line observation, we confirm for the first time the rotationally flattened hot toroid in the central <0.1 pc region, which has a rotational axis perpendicular to its geometrical major axis.

In addition, we present the observations of $NH_3$, $^{13}CS$, and $CH_3CN$ with $\sim 1''$ resolution, and follow the dynamics of the molecular accretion flow from the 0.3 pc radius to the inner 0.03 pc radius. With reference to the rotational axis of the hot toroid, we measure the rotational velocity from the molecular emission in the region. The results are consistent with an envelope with a rapid decrease of the specific angular momentum from the outer to the inner region. These new results improve the current understanding of the molecular accretion flow in an ultracompact (UC) HII region created by the embedded O-type cluster.

*Subject headings:* stars: formation — ISM: evolution — ISM: individual (G10.6–0.4)

## 1. Introduction

Rotational motions are commonly observed in massive star forming regions (see Cesaroni et al. 2007 for a general review). Growing observational evidence shows that rotating flattened core or disk–like structure ($10^1$–$10^2$ $M_\odot$, 1000–8000 AU) and massive molecular outflows (Zhang et al. 2001, 2005; Beuther et al. 2002) are commonly present in B-type star forming regions, suggesting that disks are closely related to the collapse of the molecular core. The estimated mass infall rates in these disks range from $10^{-3}$ $M_\odot yr^{-1}$ to $10^{-5}$ $M_\odot yr^{-1}$, which are comparable to or much higher than the necessary mass accretion rates to form B-type stars (Zhang 2005 and references therein). The well–studied examples of the O-type star forming regions, however, are limited, and the properties of their rotational structures are less clear. Rotational signatures have been reported toward several sources such as G24.78+0.08, G28.20–0.05, G31.41+0.31, G10.6–0.4, G16.59–0.05, G23.01–0.41, G28.87+0.07 and G20.08–0.14n (Beltrán et al. 2004, 2005; Sollins et al. 2005a, 2005b; Keto, Ho and Haschick 1987, 1988; Sollins and Ho 2005; Furuya et al. 2008; Klaassen et al. 2009; Galvan–Madrid et al. 2009). Owing to the complicated environments and their large distances, the properties of these rotating structures around the O-type star forming regions are not well investigated. In our present series of experiments, we start with the source G10.6–0.4 and establish a large molecular database with high sensitivity and high resolution (<1″, 0.03 pc). These lines probe the physical conditions in *zones* with different excitation conditions. Our goals are to unveil the morphologies and to resolve the kinematical structures of these massive toroids.

The ultracompact (UC) HII region G10.6–0.4 is one of the best studied OB cluster forming regions. The associated infrared source IRAS 18075–1956 has a luminosity of $9.2\times10^5$ $L_\odot$ (Casoli et al. 1986). From the observations of the 20 cm continuum emissions, Ho, Klein, and Haschick (1986) detected the presence of OB stars in a ∼10 pc region, and suggested that a burst of massive star formation occupies the central parsec region. The follow-up high resolution 6 cm and 1.3 cm continuum observations further suggest a concentration of OB stars in the central ∼0.1 pc region, which ionizes a bright (>2.6 Jy in 1.3 cm) UC HII region (Ho and Haschick 1986; Keto, Ho, and Haschick 1987, 1988; Sollins et al. 2005a; Sollins and Ho 2005). The detections of bright $H_2O$ and OH masers (Ho et al. 1983; Hofner and Churchwell 1996; Fish et al. 2005; Lekht et al. 2006 ) around the peak of the centimeter continuum also suggest that this is an active site of massive star formation.

The early low resolution (>5″) observations detect molecular lines in emission ($NH_3$: Ho and Haschick 1986; Keto, Ho, and Haschick 1987; CS: Omodaka et al. 1992; $C^{18}O$: Ho, Terebey, and Turner 1994) and reveal a flattened rotating dense clump (G10-main hereafter) with a velocity gradient of ∼10 $\mathrm{kms^{-1}pc^{-1}}$ at a half parsec scale radius. High–resolution (<1″) Very Large Array (VLA) observations on molecular lines have been carried out as well. However, limited by sensitivities, all previous high–resolution VLA observations (Keto, Ho and Haschick 1988; Guilloteau et al. 1988; Sollins et al. 2005; Sollins and Ho 2005) only see the absorption lines against the bright UC HII region, of which the detections are limited by the background illumination. For example, Keto, Ho, and Haschick (1988), Sollins et al. (2005a) and Sollins and Ho (2005) resolve the velocity gradient to be ∼90 $\mathrm{kms^{-1}pc^{-1}}$ in the central 2″ (0.06 pc) region. Based on these observational results of the $NH_3$ (1,1), (3,3) inversion transitions and the radiative transfer simulations, Keto (1990) suggested that the gas dynamics in G10-main are consistent with a spin-up rotational motion. However, without the high resolution observations in emission lines, the geometry of the entire region is poorly constrained. The rotational axis is therefore uncertain, which potentially leads to the systematic bias in the derived velocity gradients. Also, how does the molecular gas flow continue from the parsec scale molecular clump to the central 0.1 pc scale region was not resolved yet.

In this paper, we report our high angular resolution and high sensitivity interferometric observations of various molecular transitions toward G10-main. These 1 mm observations are sensitive to high excitation lines which trace the gas close to the central stars. We constrain the location and the orientation of the rotational motions from our highest resolution (<1″) observations of the hot core type tracers $O^{13}CS$, OCS, and $SO_2$ molecular transitions. The emissions of these transitions are only significantly detected in a <3″ region around the UC HII region, and are therefore less confused by the dynamics of the complicated and more extended features. We call the rotating structure traced by the observed $SO_2$, OCS, and $O^{13}CS$ emission the *hot toroid* hereafter. The size scale of the hot toroid is directly constrained by the detection of the selected hot core type tracers.

We then measure the velocity gradients traced by the stable molecules $CH_3CN$, $^{13}CS$, $C^{34}S$, and $NH_3$ in the plane of the hot toroid. By utilizing the excitation conditions of these molecules, we constrain the rotational velocities from 0.6 pc to the central 0.1 pc scale. Combining with the previous lower resolution CS (1-0) (Omodaka et al. 1992) and $NH_3$ (1,1) inversion line (Ho and Haschick 1986; Keto, Ho, and Haschick 1987) observations which measure the velocity gradients over a more extended region, we track the loss of angular momentum continuously over a factor of ∼50 in sizescale. In addition, we resolve the density concentration and the fragmentation signatures in the central 0.5 pc region. Details about the observations, and the data reductions are introduced in Section 2. The results and the method to analyze the rotational signature are presented in Section 3.

## 2. Observation and Data Reduction

We provide the detailed information of our observations, and the data reductions. The properties of the molecular line data are listed in Table 1. The observational parameters of the selected molecules are summarized in Table 2.

### 2.1. Line Observations

#### *2.1.1. The $CH_3CN$ Observations*

We observed the $CH_3CN$ J=12-11 transitions using the Submillimeter Array (SMA)[1] in the extended configuration on 2005 September 05. For more on the SMA and its specifications, see Ho, Moran & Lo (2004). The frequency resolution in that observation is 390 kHz (∼0.53 $kms^{-1}$). The basic calibrations of these data are done in Miriad, and additional self-calibration on the source are done in Miriad, and imaging is done in AIPS.

In this paper, we present the $CH_3CN$ 12(5)-11(5), and 12(6)-11(6) mapping results to trace the high temperature (>250 K) and high density gas, observed to be concentrated to a small region close to the UC HII region (for more analysis of these $CH_3CN$ data see Klaassen et al. 2009). Continuum emissions are averaged from the line-free channels and then subtracted from the line data. To enhance the signal–to–noise ratio, we average the visibilities of these two lines to form a synthesized line ($CH_3CN$ $K_{56}$ line hereafter; these two transitions have upper energy levels of 242 K and 319 K, respectively). The signal to noise ratio will be enhanced by a factor of $\sqrt{2}$ if both the 12(5)-11(5) and the 12(6)-11(6) lines are detected. However, in regions where the temperature is not high enough to excite the 12(6)-11(6) transition, the signal to noise ratio will be degraded by a factor of $\sqrt{2}$. Thus we argue that the temperature traced by the $K_{56}$ line is biased to be higher than 300 K. We find that the emission of the $K_{56}$ line is mainly contributed by the gas distributed within a 0.1 pc region of the UC HII region.

---

[1] The Submillimeter Array is a joint project between the Smithsonian Astrophysical Observatory and the Academia Sinica Institute of Astronomy and Astrophysics, and is funded by the Smithsonian Institution and the Academia Sinica.

The $CH_3CN$ line is imaged using natural weighting, which leads to the synthesized beam of $1''.5\times1''.1$ with position angle $56^o$. The observed rms noise for the channel maps is 0.12 Jy beam$^{-1}$.

#### 2.1.2. The $^{13}CS/C^{34}S$ (5–4), $OCS/O^{13}CS$ (19–18) and $SO_2$ Observations

We observed the $^{13}CS$ (5–4), $OCS/O^{13}CS$ (19–18), $SO_2$ 5(2,4)–4(1,3) and $SO_2$ 18(1,17)–18(0,18) transitions toward G10.6-0.4 using the SMA in the compact configuration and the very extended configuration on 2009 June 10 and 2009 July 12, respectively. We also observed $^{13}CS$ (5–4) in the subcompact configuration on 2009 January 31 . The frequency resolutions in these observations are 390 kHz ($\sim$0.53 kms$^{-1}$).

The basic calibrations are done in Miriad, while the self-calibration and imaging of these data are done in AIPS. Continuum emissions are averaged from the line-free channels and then subtracted from the line data. We combined the $^{13}CS$ data of all three array configurations, which yields a synthesized beam of $1''.3\times0''.66$ with a position angle $79^o$. The observed rms noise of the $^{13}CS$ (5-4) line channel maps is 0.05 Jy beam$^{-1}$. We combined the compact array and the very extended array data for the other molecules. To better constrain the geometry and velocity gradient in the 0.1 pc scale rotating core, we imaged the OCS and $SO_2$ lines using uniform weighting, yielding the highest angular resolution of $0''.62\times0''.43$ with a position angle $22^o$. We imaged the weaker $O^{13}CS$ line using natural weighting to improve the signal–to–noise ratio. The $C^{34}S$ line is imaged using *robust=0* for optimizing the resolution and sensitivity.

#### 2.1.3. The $NH_3$ (3,3) Main Hyperfine Line Observations

We observed the $NH_3$ (3,3) main hyperfine line toward G10.6–0.4 ($\alpha$(J2000) $=18^h10^m28^s.683$, $\delta$(J2000) $= -19^o55'49''.07$) using the NRAO[2] VLA in the C-configuration on 2009 July 27. The frequency resolution in these observations is 97.656 kHz ($\sim$1.23 kms$^{-1}$). We observed 3C286, 0319+415, and 1820–254 as our absolute flux, passband, and gain calibrators.

The basic calibrations, self-calibration, and imaging of these data are done in AIPS. The size of the synthesized beam is $1''.8\times1''.2$ with a position angle $0.96^o$, and the rms noise is 8 mJy beam$^{-1}$ ($\sim$9 K). Continuum emissions are averaged from the line-free channels and then subtracted from the line data. Our data are consistent with the earlier $NH_3$ (3,3) D-array data reported in Keto,

[2]The National Radio Astronomy Observatory is a facility of the National Science Foundation operated under cooperative agreement by Associated Universities, Inc.

Ho, and Haschick (1987) when smoothed to the same resolution. However, our higher angular and velocity resolutions provide better constraints for the kinematics in the inner 10″ (0.3 pc) region. For the $NH_3$ kinematics in the more extended (and cooler) region, we will directly quote the $NH_3$ (1,1) main hyperfine inversion transition result from Keto, Ho and Haschick (1987).

### 2.2. Continuum Observations

We construct the continuum band at 1.3 mm from the line–free channels in the SMA subcompact, compact, and very–extended array data mentioned in Section 2.1.2. We combined the data of all three array configurations, which yields the synthesized beam of $0''.79\times0''.58$ with position angle $60^o$. With the large 4 GHz total bandwidth, a low rms noise of 3 $\mathrm{mJy\,beam^{-1}}$ is achieved.

We also retrieved and reprocessed the 1.3 cm VLA A-array continuum data from the data archive. The synthesized beam of this 1.3 cm continuum map is $0''.12\times0''.07$ with position angle $58^o$. See Sollins et al. (2005) and Sollins and Ho (2005) for the detailed descriptions of these VLA continuum observations.

## 3. Results

### 3.1. The Continuum Emissions

Figure 1 shows the 1.3 mm and the 1.3 cm continuum emission maps. The comparison of the continuum emissions at these two frequencies suggests that

- Inwards of the $\sim3''$ region, the 1.3 mm continuum is contributed by the free-free emission of the UC HII region and the thermal dust emissions.
- The 1.3 mm continuum is dominantly contributed by the thermal dust emission in the extended region.
- The extended dust emissions show clumpy and protrusive morphology.

The measured 1.3 mm continuum flux is about 10 Jy within a 10″ (0.3 pc) region, and about 6 Jy within the central 4″ region. By subtracting the measured 1.3 mm continuum flux with the predicted lower limit of free-free continuum flux at 1.3 mm (Keto, Zhang, and Kurtz 2008), we find that the upper limit of the thermal dust emission at 1.3 mm within the central 4″ region is 2.7 Jy. Assuming an average temperature of 200 K and assuming a gas-to-dust mass ratio of 100, the 2.7 Jy flux corresponds to the $H_2$ mass of 140, 230 and 380 $M_\odot$ for $\beta$= 1, 1.5 and 2, respectively (Lis et al. 1998). Assuming an average temperature of 50 K, without including the central 4″ region, the $H_2$ mass estimated by dust emission is 240, 400, and 660 $M_\odot$ for $\beta$= 1, 1.5 and 2, respectively. The total $H_2$ mass enclosed in the central 10″ region is therefore 240–1040 M⊙. We note that extended emissions may have some missing flux, and the mass derivations for that region should be treated as the lower limit.

### 3.2. The OCS, $O^{13}CS$, $SO_2$ and $CH_3CN$ Emissions —The Hot Toroid

Figure 2 shows the moment zero, one and two[3] maps of the OCS (19–18) line, the $O^{13}CS$ (19–18) line, the $SO_2$ 18(1,17)–18(0,18) and 5(2,4)–4(1,3) lines, and the $CH_3CN$ $K_{56}$ line. From these maps, first we can see that these transitions are only detected in a small (2″-3″) region around the UC HII region. From the moment 2 and moment 1 maps, we see that the local velocity dispersion is generally small as compared with the velocity range traced by each molecule, suggesting that the organized dynamics dominates over the thermal/turbulent motions. A prominent common dynamics feature detected by all these tracers is a velocity gradient from southeast to northwest along the major axis of the flattened structure, which suggests a rotating toroid. From the velocity gradient, we determined the position angle to be $140^o \pm 5^o$. The geometry of the $CH_3CN$ emission appears to be more spherical, which is due to the relatively low resolution.

Assuming that the molecular abundance is uniform in the size scale of the hot toroid, the ratio of the $O^{13}CS$ intensity to the OCS intensity shows the relative distribution of the column density. Figure 3 shows the ratio of the velocity integrated flux density of the $O^{13}CS$ and the OCS lines[4]. From this figure we can see that the distribution of the dense molecular gas is consistent with a $\sim 2'' \times 1''$ toroid aligned along the southeast-northwest axis, and the $O^{13}CS$/OCS ratio is about 0.2 in that region. By comparing the $O^{13}CS$/OCS ratio maps with the high resolution $NH_3$ (3,3) satellite hyperfine line absorption optical depth maps (Figure 3, see Sollins et al. 2005a for the introduction of the high resolution $NH_3$ data), we can see that the dense molecular gases traced by these two molecules are in excellent agreement with each other, suggesting that these molecular line transitions are indeed tracing the hot and dense rotating toroid in the most central region. In this particular source, the UC HII region is optically thick in the centimeter band (Keto, Ho and Haschick 1987, 1988). We note that in the millimeter band, both the UC HII region and dust are optically thin, which allows us to probe the gas motions at the back or the far side of the UC HII region, and therefore improves the constraint on the gas kinematics as compared to previous centimeter band experiments (Keto, Ho and Haschick 1987, 1988; Sollins and Ho 2005). Assuming that the abundance ratio [OCS]/[$O^{13}CS$] is 30 (Gibb et al. 2000), we obtain the average optical depth of the $O^{13}CS$ and the OCS J=19–18 transitions in the region with $>3\sigma$ detections to be 0.17 and 5.12, respectively. The detected brightness temperatures of the selected $SO_2$ lines are comparable to the brightness temperature of the OCS line. Therefore, we suggest that these lines

---

[3]The moment two is estimated from $\frac{\int v^2 F_v dv}{\int F_v dv}$.

[4]When generating this ratio map, both the OCS and the $O^{13}CS$ (19-18) lines are imaged with natural weighting.

have a similar beam filling factor, and are generally optically thick.

It should be noted that our high resolution observations of these hot core and shock tracers suggest that the molecular gas in the central 3″ (0.09 pc) region is fragmented and may be interacting with the ionized gas and the outflows. In this paper we focus on the discussion of the rotational signatures shown in these molecular line observations, which are specifically associated with the strong gravitational source, an OB cluster at the center. Discussions of the high velocity molecular outflows and shock signatures will be presented in a separate publication.

## 3.3. The Structure of Specific Angular Momentum

The molecules $^{13}CS/C^{34}S$, $NH_3$ and $CH_3CN$ have relatively stable abundances as compared to the other hot core and shock tracers. The emission lines of these molecules trace a broad range of excitation conditions. By comparing the rotational signatures traced by these molecules, we successfully follow the rotational signature from a parsec scale radius to the central 0.03 pc radius around the OB cluster. We introduce the terminal velocity method to analyze the rotational velocity as a function of radius and show the decrease of specific angular momentum inward of the 0.3 pc radius.

### *3.3.1. The Spatial Distribution of Tracers*

Figure 4 shows the moment zero map of the $^{13}CS$ (5-4) line, the $NH_3$ (3,3) main hyperfine inversion line, the $CH_3CN$ $K_{56}$ line, and the 1.3 cm continuum emission (see Sollins et al. 2005 for the introduction of the 1.3 cm continuum observation.) Our high resolution observations of the $NH_3$ (3,3) emissions show complicated morphology with many protrusions in an extended (∼20″, 0.6 pc) region. The distribution of the dense gas tracer $^{13}CS$ (5-4) is consistent with a highly non-uniform flattened molecular clump, and may harbor more than one gravitationally bound dense core; the distribution and kinematics of the $C^{34}S$ (5-4) are consistent with $^{13}CS$ (5-4). The velocity integrated flux of $C^{34}S$ (5-4) is about 1.8 times the velocity integrated flux of the $^{13}CS$ (5-4). Assuming the abundance ratio of $[^{13}CS]/[C^{34}S]$ is 0.4 (Frerking et al. 1980), the $[CS]/[C^{34}S]$ ratio is 22.6 (Omodaka et al. 1992), the measured $C^{34}S$ (5-4) optical depth ranges from ∼0.05 to ∼2 over most of the regions with $>3\sigma$ detections of $^{13}CS$ (5-4), and the $C^{34}S$ optical depth can be higher than 4 in the densest (∼1″ scale) regions. These results suggest that the $^{13}CS$ (5-4) line is at least marginally optically thin in most regions, and therefore is a good tracer of the gas distributions and gas kinematics. In this paper we only show the images of the optically thinner $^{13}CS$ isotopologues to trace the detailed structures and the kinematics in the inner 0.5 pc clump. The high temperature and high density tracer $CH_3CN$ $K_{56}$ line is only detected in a 3″ (0.09 pc) region around the UC HII region, which allows us to follow the kinematics of the hot molecular gas directly associated with the UC HII region.

#### *3.3.2. The Position–Velocity Diagram*

Before the systematic analysis, we first visually inspect the local structures and the rotational signatures from the position–velocity (PV) diagrams. Figure 5 compares the PV diagram of the $NH_3$ main hyperfine emission line, the $^{13}CS$ (5-4) line, and the $CH_3CN$ $K_{56}$ line. The PV cuts are centered at the coordinates of R.A. = $18^h10^m28.64^s$ and decl. = -$19^o55'49.22''$ with position angle P.A. = $140^o$ From this figure, we can clearly identify the fast rotational signatures as traced by each molecule; globally, the molecular gas tends to be more redshifted as the angular offset becomes more negative (or less positive). In addition, we found that the motions are consistent with an inward decrease of the specific angular momentum ($\vec{r}\times\vec{v}$). The $CH_3CN$ traces the fast rotating hot toroid, with the rotational velocity gradient of ∼100 kms$^{-1}$pc$^{-1}$ at a 0.03 pc radius. In the $^{13}CS$ (5-4) PV diagram (Figure 5), we observed very prominent rotational signatures from angular offset -1$''$ to 4$''$, with an enhanced and lumpy distribution. The velocity gradient of that rotational signature is ∼60 kms$^{-1}$pc$^{-1}$ at a 0.08 pc radius. This rotational motion can be gravitationally bound by 430 $M_\odot$. By comparing the rotation curve traced by $^{13}CS$ and $CH_3CN$ in the pv diagrams (Figure 5, right), we obviously see a spin-up rotational signature. Note the fast rotating signature traced by the $CH_3CN$ $K_{56}$ line should also contribute in the $^{13}CS$ PV diagram. However, it is blended with the outer (0.08 pc radius) slower rotating signature. If we use methods such as fitting 2D gaussian to extract the velocity gradient of the hot toroid from the $^{13}CS$ PV diagram, the confusion of the structure at outer radii leads to an underestimation of the velocity gradient, and also leads to the inconsistency with velocity gradient estimated from the $CH_3CN$ data or other hot core tracer data. The $^{13}CS$ rotational signature at the 0.08 pc radius is less confused by the kinematics in larger radius since this structure is the dominant feature in that particular angular offset and velocity range.

The line width of the $NH_3$ (3,3) main hyperfine line is broad (generally >1 kms$^{-1}$, and >>1 kms$^{-1}$ for low intensity emissions estimated from the moment two map) over the map, but the lines still show the general southeast to northwest velocity gradient of ∼15 kms$^{-1}$pc$^{-1}$ at 0.3 pc radius. The observed broad linewidth emissions of $^{13}CS$ and $NH_3$ suggest that the molecular core is interacting with outflow or the UC HII regions.

From these PV diagrams, we select the $CH_3CN$ data in the inner 0.05 pc radius, the $^{13}CS$ (5-4) data in the inner 0.2 pc radius, and the $NH_3$ (3,3) main hyperfine inversion line in the 0.1–0.3 pc radius to perform the terminal velocity analysis (Section 3.3.3). The selected range of radii ensures each molecule to have greater than $3\sigma$ detection.

#### *3.3.3. The Analysis of Terminal Velocity and the Velocity Gradient*

The determination of the velocity gradients of the hot core tracers $O^{13}CS$, OCS, $SO_2$, and $CH_3CN$ can be done via Gaussian or other linear fittings. However, the velocity gradients for the $^{13}CS$, $C^{34}S$ and $NH_3$ (3,3) emission structures are less straightforward to estimate. These molecules are present over a larger region, with rich resolved local density structures and large local velocity dispersions. Any fitting procedure will be biased by the underlying structures in mass, excitation, or velocity.

To sketch the rotation curve in the massive cluster forming regions, we apply the concept of the terminal–velocity method used in extragalactic studies (a review of commonly applied methods for measuring the galaxy rotation curve is given by Sofue and Rubin 2001). Using this method to extract the rotational velocity has the advantage over other fitting procedure since the systematic bias in this method is visually trackable and understandable. From the observed geometry (Section 3.2) of the hot toroid, we assume the system is edge–on. The terminal velocity $v_t$ is then defined by a velocity at which the intensity $I$ equals to

$$I_t = \eta I_{max},$$

and the rotational velocity is defined by

$$v_{rot} = v_t - v_{lsr},$$

where $v_{lsr}$ = -3 kms$^{-1}$ for G10.6–0.4, and $\eta$ is a free parameter ranging between 0 and 1. In G10.6-0.4, the dynamics in the inner and the outer regions are severely blended along the line–of–sight, and the centroid velocity at each radius can be confused by the local dense structures, In such case, the systematic offset of $v_{rot}$ from the real rotation curve is larger than the uncertainties contributed by the thermal or local turbulent motion (0.5–1 kms$^{-1}$ effect).

Before performing the terminal–velocity method, we smooth each image with $1''.3$ kernal to suppress the small–scale fluctuations caused by local intensity peaks. To conservatively demonstrate the decrease of the specific angular momentum with respect to the values in the outer radius reported in the previous low–resolution observations, we purposely choose $\eta = 0.2$, which systematically provide higher estimation of rotational velocity. Instead of explicitly showing the error bar, we present the $\eta = 0.2$ and $\eta = 0.8$ results, of which the comparison gives the sense of uncertainties involved. For the $\eta = 0.8$ case, if at a certain radius the intensity is below $\eta I_{max}$ in the entire velocity range, we define $v_{rot}$ at that radius to be the velocity at the peak of the intensity spectrum. We demonstrate the robustness of the analysis from the consistency of rotational velocities traced by different molecules. We define the center of the system at the location where the centroid velocity of the $CH_3CN$ equals to $v_{lsr}$, and evaluate the velocity gradient at each radius r by $v_{rot}/r$. The results of the terminal velocity analysis is provided in the next section.

The examples of the $^{13}$CS (5-4) terminal–velocities estimated using $\eta$ = 0.2 and $\eta$ =0.8 are shown in Figure 6, with the overlay of the pv diagram. From this figure, we see that choosing $\eta$ =0.8 provides significantly lower terminal–velocity. We note that if the molecular gas in an edge–on system rotates concentrically, for a specific line–of–sight, its projected spatial offset is equal to the inner most radius intersecting with that line–of–sight. The motion of the molecular gas at that inner most radius is parallel to the line–of–sight, and the redshift of the emission directly reflects the rotational velocity at that certain radius. In a spin–up system, at a certain spatial offset, the higher velocity emission is likely to be contributed from the gas in the inner radius. Hence, using $\eta = 0.2$ should better reflect the motions at the inner most radius.

#### 3.3.4. *The Velocity–Gradient and the Specific Angular Momentum*

In Figure 7, we plot the velocity gradients in the axis with position angle of $140^o$, traced by $CH_3CN$, $^{13}CS$ (5-4), and $NH_3$ (the 2009 C-array data and the D-array data from Ho and Haschick 1986, Keto, Ho, and Haschick 1987), CS (Omodaka et al. 1992) at different radii. The local scattering of the data points reflects the statistical uncertainties in the measurements. According to the measured velocity gradients, we estimate the average specific angular momentum as a function of radius, and summarize the derived results in Figure 8.

From these diagrams, we see that in zones where the rotational velocity can be measured by more than one molecular line, the results from different molecular lines are consistent. This indicates that applying the same $\eta$ value for different molecular lines is robust. Additionally we see that the estimations with $\eta = 0.2$ and $\eta = 0.8$ are qualitatively consistent with each other. The $\eta = 0.8$ results show lower velocity gradient and lower specific angular momentum. This is because of the bias by the mass along the line–of–sight. The results show that from the 1.5 pc radius to the 0.3 pc radius, the observed specific angular momentum of the molecular gas is nearly conserved, or has a slight decrease. Inside the 0.3 pc ($10''$) radius, the specific angular momentum decreases rapidly, from ~1.5 pc·km·s$^{-1}$ (at 0.3 pc radius) to <0.1 pc·km·s$^{-1}$ (at 0.03 pc radius). From the results, we find a turnover radius $r_T$ of about 0.3 pc within which specific angular momentum experiences a significant loss. For $r<r_T$, we find a specific angular momentum loss rate of ~5.2 pc·km·s$^{-1}$ per parsec.

We note that the diagnostic model in Keto (1990) suggested that the gas flow does not conserve angular momentum on any scale. However, that previous model was constructed based on the observations of the $NH_3$ line emissions with $\sim 10''$ (0.3 pc) resolution, and the observations of the $NH_3$ line absorption with $\sim 0''.3$ (0.009 pc) resolution. Inward of the 0.3 pc radius, the model had been uncertain owing to insufficient sampling of the spatial scales. Our new observations presented in this paper can be viewed as a more detailed demonstration of the decreasing specific angular momentum suggested in the previous model.

In addition to the terminal–velocity method, we also make alternative estimations of the velocity gradient by referencing to intensity–weighted mean velocities. We perform the linear regression fittings to obtain the slopes of the intensity–weighted mean velocity, which can be interpreted as velocity gradients. The results are summarized in Table 3. When fitting the $^{13}CS$ velocity, we exclude the emission enclosed by the dashed box in Figure 5 since the velocity of the brightest component in that region is close to the systemic velocity of the entire molecular cloud (-3 kms$^{-1}$), will be dominated by that motion. The extended emission in that region may be the corresponding outflow. The results are shown in Table 3. For the $CH_3CN$ line, since the scale of the synthesized beam is comparable to the scale of the emission region, we perform linear regression fittings of the gradients of intensity-weighted mean angular offset along the velocity axis. The emission at angular offset $\sim 0''$ with velocity >2 $kms^{-1}$ does not have a high signal–to–noise ratio. Thus it is not clear whether it is a real structure. We therefore provide two fitting results; the lower velocity gradient value is obtained by excluding the data with velocity >2 $kms^{-1}$.

The observed velocity gradients of the OCS (19-18) line, the $O^{13}CS$ (19-18) line, the $SO_2$ 18(1,17)-18(0,18) and 5(2,4)-4(1,3) lines, and the $CH_3CN$ lines are consistent.The errors take into account the uncertainties introduced by the local velocity dispersions at both ends of the rotational signatures. These values are also consistent with the velocity gradient of the $CH_3CN$ line obtained from the terminal–velocity method (Figure 7). Assuming that the hot and dense core is edge on, the observed velocity gradients can be gravitationally bound by $\sim$200 $M_\odot$. However, from the Lyman continuum emission rate, Sollins et al. (2005) estimated the stellar mass embedded in the UC HII region to be $\sim$175 $M_\odot$, which dominates the enclosed mass estimated from the rotation curves. We therefore suggest that the geometry of the hot and dense molecular gas in the central $3''$ (0.09 pc) region is a rotating ring/torus which encircles the UC HII region. The velocity gradient of the $NH_3$ line obtained from the linear regression fitting is significantly lower than those values obtained from the terminal–velocity method with $\eta$=0.2, however, consistently suggests the low specific angular momentum inward of the 0.3 pc radius.

The comparison of the $CH_3CN$ distribution with the $^{13}CS$ distribution (Figure 4) suggests that the rotational centers of the hot toroid and the lumpy $^{13}CS$ ring/toroid are offset by a few arcseconds. The peak velocities of $CH_3CN$ and $^{13}CS$ are also offset by a few $kms^{-1}$ from the overall systemic velocity of -3 $kms^{-1}$. These offsets can be explained by fragmentation within the central core, resulting in peculiar motions, as well as opacity/excitation and chemical effects which may bias the peak emission velocities. Moreover, the motions within the central 0.3 pc core appear quite turbulent with velocity dispersions >5 $kms^{-1}$. At smaller scales (<0.1 pc radius), strong magnetic braking by a few mG magnetic field may efficiently remove the specific angular momentum (see Keto, Ho & Haschick 1987, and Girart et al. 2009 for derivations). Between 0.1 and 0.3 pc scale, it is likely that a large fraction of the specific angular momentum remains in the orbital motions. This results in reduced specific angular momentum at the scale of the fragments which are able to form and which are detected. Note that the resolved $NH_3$ protrusions and the large velocity dispersion of the $^{13}CS$ (5-4) also suggest that the outflow, shock, and the induced turbulence may be closely related to the transfer of angular momentum. The relative importance of these processes, and other possible ones, remains to be elucidated by future observations.

### 3.4. Mass Estimates Using Molecular Lines

#### *3.4.1. The Hot Toroid*

We estimate the observed molecular mass of the hot toroid from the integrated intensities of the $CH_3CN$ J=12–11, K=5, 6, and the OCS, $O^{13}CS$ J=19–18 emissions. We suggest that the hot molecular gas with an average temperature of T$\sim$300 K (see also Klaassen et al. 2009: 323$\pm$105 K) dominates the contribution to the flux in these transitions. By adopting the $CH_3CN$ abundance $[CH_3CN]/[H_2]=10^{-8}$–$10^{-7}$ (Kalenskii et al. 2000; Klaassen et al. 2009), and assuming $CH_3CN$ J=12, K=5, 6 transitions are optically thin, which is typically true in hot molecular cores, the derived $H_2$ mass from K=5 lines in the central 0.06 pc region is 12–1.2 $M_\odot$, and the derived $H_2$ mass from K=6 lines in the central 0.06 pc region is 7.2–0.72 $M_\odot$. The deviation of the derived $H_2$ mass from the two transitions may be caused by the error in our temperature assumption.

Assuming that the $O^{13}CS$ J=19–18 is generally optically thin, we estimate the lower limit for the $H_2$ mass as traced by $O^{13}CS$ to be 15–15000 $M_\odot$ for the range of the abundance ratio $[OCS]/[H_2]$ of $10^{-6}$–$10^{-9}$ (see Viti et al. 2004 and references therein for the abundance ratio). Since the hot toroid is interacting with the central OB cluster, the ionized gas, and the outflows, we expect a highly enhanced OCS abundance in the detected region, and the molecular mass of the hot toroid should be closer to the lower estimate of 15 $M_\odot$.

By observing the $NH_3$ (3,3) inversion transition, Keto, Ho, and Haschick (1988) estimated the $H_2$ mass in the similar region to be 3–60 $M_\odot$ using the assumption of abundance $[NH_3]/[H_2]=10^{-6}$–$10^{-7}$ (Ho and Townes 1983). Considering the uncertainties in the $CN_3CN$, OCS and $NH_3$ abundance and the foreground contribution to the $NH_3$ (3,3) line, the mass estimates by these three molecules are consistent. By comparing the binding mass of the $^{13}CS$ motion (at 0.08 pc radius: 300 $M_\odot$) with the mass of the OB cluster (175 $M_\odot$) and the mass of the hot toroid traced by the $CH_3CN$ line, we argue that the bulk of the $^{13}CS$ dense lumpy ring/toroid is encircling the hot toroid and the OB cluster. The observed velocities are consistent with gravitationally bound motions due to the enclosed mass at each radius.

#### *3.4.2. The Larger Scale Mass*

Assuming that the abundance ratio $[CS]/[H_2]$ is $10^{-6}$–$10^{-9}$ (Viti et al. 2004), an average temperature of 50 K, and assuming optically thin emission, we estimate the lower limit of the $H_2$ mass traced by the $^{13}CS$ (5-4) to be $10^2$–$10^5$ $M_\odot$. By adopting the abundance ratio $[CS]/[H_2]=10^{-9}$, Omodaka et al. (1992) estimate the $H_2$ mass of the entire clump traced by $C^{34}S$/CS (2-1) to be $4{\cdot}10^4$ $M_\odot$, which agrees well with our estimate. Using $C^{18}O$ (2-1), Ho, Terebey and Turner (1994) estimated the $H_2$ mass in the central 30$''$ (0.9 pc) region to be 2500 $M_\odot$, suggesting that the average value of [CS]/[$H_2$][5] is close to $10^{-7}$–$10^{-8}$. We note that from this large scale (∼1pc) core, about 10% of the mass has condensed and formed the visible massive stars, and only 10% of the stellar mass is in a remnant hot toroid. The star formation efficiency is very high.

#### *3.4.3. The Centrifugal Velocities*

In Figure 7, we summarize the required velocity gradients to balance against the gravity of the enclosed molecular and stellar masses at a number of radii. Values of $NH_3$ mass, CS mass, $C^{18}O$ mass, and the embedded stellar mass are quoted from the literature (Keto, Ho and Haschick 1987; Omodaka et al. 1992; Ho, Terebey and Turner 1994; Sollins et al. 2005). By comparing the required velocity gradients with the measured velocity gradients, we suggest that there may be a crossover at about 0.3 pc radius. The dynamics of the gas in the entire region is dominated by gravity. Inward of the crossover, some mechanism removes excess angular momentum allowing the flow to maintain a high inward velocity and high accretion rate even as the flow progresses to smaller and smaller radii. Note that the location of this crossover is consistent with the turnover radius of the specific angular momentum (Figure 8). The formation of a flattened structure with spin-up motions within a fluffy core, as seen in Figure 4, is consistent with this scenario.

## 4. Conclusions

We have carried out arcsecond resolution SMA and VLA/eVLA observations in the massive cluster forming region G10.6–0.4. The main results are as follows:

- By comparing the $O^{13}CS$/OCS ratio with the $NH_3$ (3,3) main and satellite hyperfine line ratio, we confirmed a $\sim 2'' \times 1''$ rotationally flattened dense toroid around the UC HII region. Owing to the uncertainties in abundance, the molecular mass of the dense toroid can be from a few to a few tens of solar mass. The dense toroid is marginally rotationally supported against the gravity of the encircled stellar cluster of ∼175 $M_\odot$.
- Using the terminal–velocity method, we follow the rotational velocity from the 0.3 pc scale radius to the 0.03 pc scale radius. The results suggest a rapid decrease of specific angular momentum inward of the 0.3 pc radius, with a mean specific angular momentum loss rate of ∼5.2 pc·$kms^{-1}$ per parsec.
- We detect local structures in the $^{13}CS$ (5-4) observations, suggesting that the system is clumpy and self-gravitating. The fragmentation may account for 25%–40% of the loss of specific angular momentum.

[5]The [CS]/[$H_2$] ratio in Orion hot core, compact ridge, and extended ridge is about $10^{-8}$ (Sutton et al. 1995; Charnley 1997).

Baobab Liu thanks Meri Stanley and Mark Claussen for help in preparing the VLA/eVLA observations. *Facilities:* SMA, VLA

| Transition | Frequency (GHz) | $E_u$ (K) | Eins. A. ($s^{-1}$) | Note |
|---|---|---|---|---|
| $SO_2$ 5(2,4)–4(1,3) | 241.615779 | 23.5 | $8.46{\cdot}10^{-5}$ | |
| $SO_2$ 18(1,17)–18(0,18) | 240.942788 | 162.0 | $7.02{\cdot}10^{-5}$ | |
| OCS 19–18 | 231.060991 | 110.2 | $3.58{\cdot}10^{-5}$ | |
| $O^{13}CS$ 19–18 | 230.317500 | 109.9 | $3.54{\cdot}10^{-5}$ | |
| $NH_3$ (3,3) | 23.870129 | 124.5 | $2.56{\cdot}10^{-7}$ | |
| $^{13}CS$ (5–4) | 231.220768 | 33.1 | $2.51{\cdot}10^{-4}$ | |
| $CH_3CN$ 12(5)–11(5) | 220.641096 | 242 | $7.6{\cdot}10^{-4}$ | $K_{56}$ line |
| $CH_3CN$ 12(6)–11(6) | 220.594438 | 319 | $6.9{\cdot}10^{-4}$ | |

Table 1: Table of the selected molecular transitions. The quantum number of the transitions are listed in the first column. Their frequencies and upper-level-energy are listed in the second and the third column. The Einstein coefficient of each transition is listed in the fourth column. The last two transitions are averaged to be the $K_{56}$ line.

| Transition | bpa | $b_{maj}$ | $b_{min}$ | uv Sampling Range (k$\lambda$) | rms in Image (JY/beam) |
|---|---|---|---|---|---|
| $SO_2$ 5(2,4)–4(1,3) | $21^o$ | $0''.62$ | $0''.43$ | 6-393 | 0.08 |
| $SO_2$ 18(1,17)–18(0,18) | $23^o$ | $0''.62$ | $0''.43$ | 6-393 | 0.08 |
| OCS 19–18 | $22^o$ | $0''.62$ | $0''.43$ | 6-393 | 0.08 |
| $O^{13}CS$ 19–18 | $73^o$ | $0''.97$ | $0''.61$ | 6-393 | 0.05 |
| $NH_3$ (3,3) | $0.96^o$ | $1''.8$ | $1''.2$ | 2.2-270 | 0.008 |
| $^{13}CS$ (5–4) | $80^o$ | $1''.3$ | $0''.66$ | 5.4-393 | 0.05 |
| $CH_3CN$ 12(5)–11(5) | $56^o$ | $1''.5$ | $1''.1$ | 26-145 | 0.12 |
| $CH_3CN$ 12(6)–11(6) | | | | | |

Table 2: The observational parameters.

| Transition | Fitted Velocity Gradient ($kms^{-1}pc^{-1}$) | Radius (pc) |
|---|---|---|
| $CH_3CN$ $K_{56}$ | $146.0^{+27}_{-20}/109.9^{+18}_{-13}$ | 0.027 |
| $SO_2$ 5(2,4)-4(1,3) | $95^{+40}_{-10}$ | |
| $SO_2$ 18(1,17)-18(0,18) | $105^{+60}_{-25}$ | |
| OCS 19-18 | $85^{+40}_{-20}$ | |
| $O^{13}CS$ 19-18 | $85^{+40}_{-20}$ | |
| $^{13}CS$ (5-4) | $38.2\pm2.8$ | 0.075 |
| $^{13}CS$ (5-4) | $15.6\pm1.5$ | 0.15 |
| $NH_3$ (3,3) | $8.7\pm2.2$ | 0.27 |

Table 3: The fitted velocity gradients. The velocity gradients of the hot core type tracers are evaluated in the same spatial region. See Section 3.3.4 for the details of fittings.

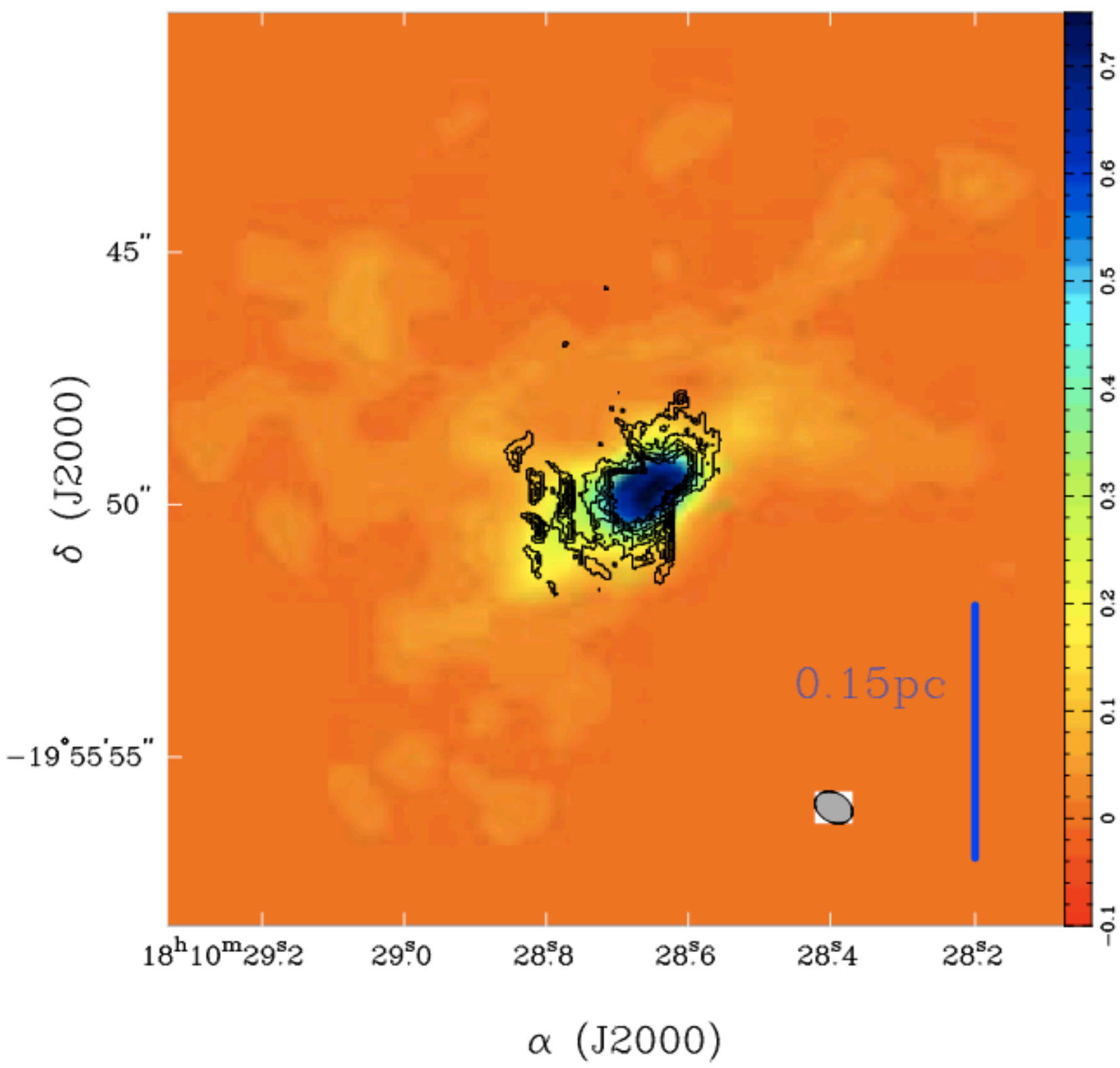

Figure 1: 1.3 mm continuum image (color) and the 13 mm continuum image (contour). The synthesized beam of the 1.3 mm continuum image is shown in bottom right of this figure. Labels of the color bar have the unit of Jy beam$^{-1}$. Contours start from 5% of the 13 mm continuum peak, with 5% steps.

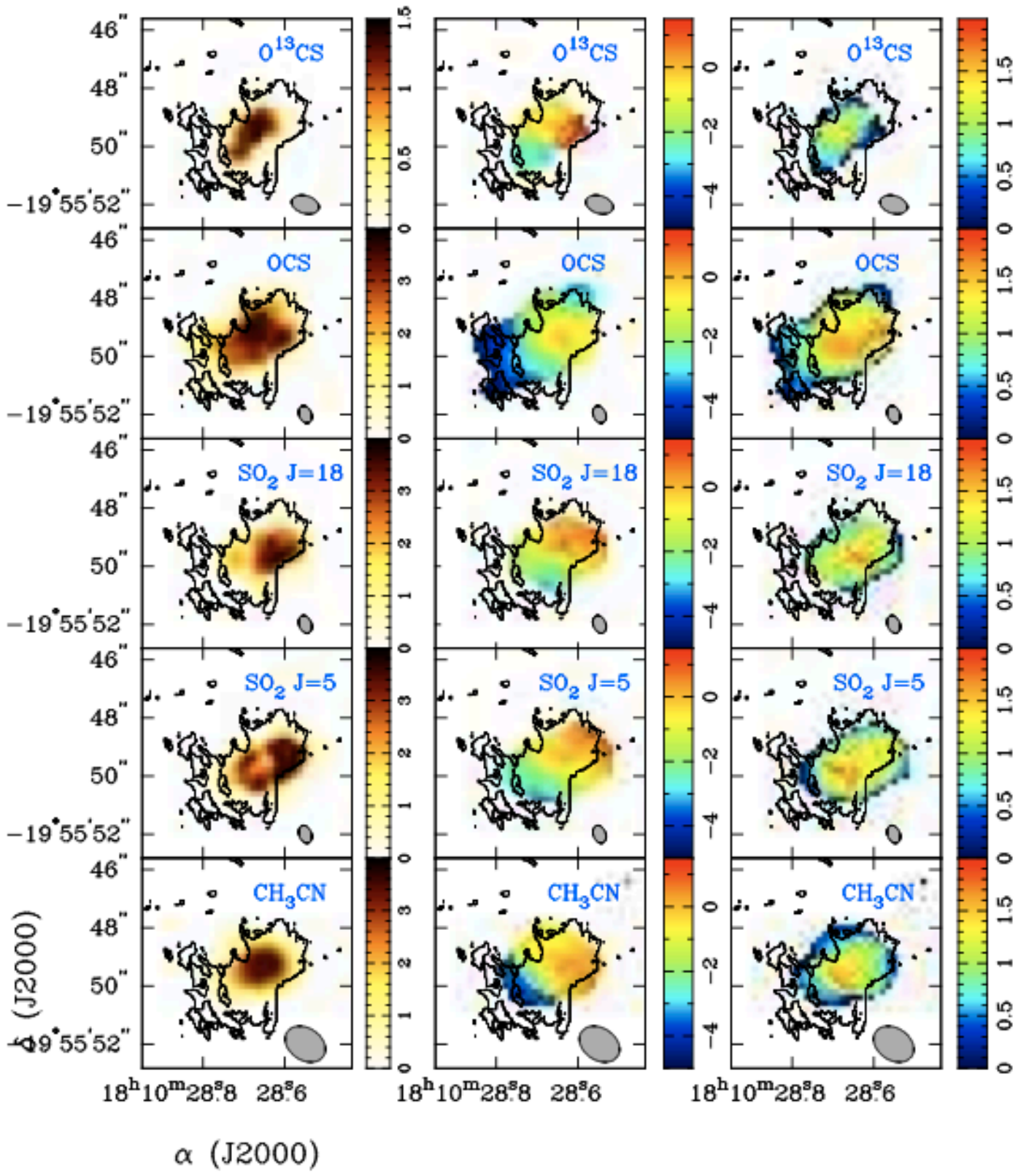

Figure 2: **Left:** Moment zero maps of the $O^{13}CS$ (19–18) line, the OCS (19–18) line, and the $SO_2$ 18(1,17)–18(0,18) and 5(2,4)–4(1,3) lines, and the $CH_3CN$ $K_{56}$ line. Contours show the 3% level of the 1.3 cm continuum emission. The unit of the color bar is (Jy beam$^{-1}$)·(m/s). **Middle:** Moment one maps of the $O^{13}CS$ (19–18) line, the OCS (19–18) line, and the $SO_2$ 18(1,17)–18(0,18) and 5(2,4)–4(1,3) lines, and the $CH_3CN$ $K_{56}$ line. Contours show the 3% level of the 1.3 cm continuum emission. The unit of the color bar is km s$^{-1}$. **Right:** Moment two maps of the $O^{13}CS$ (19–18) line, the OCS (19–18) line, and the $SO_2$ 18(1,17)–18(0,18) and 5(2,4)–4(1,3) lines, and the $CH_3CN$ $K_{56}$ line. Contours show the 3% level of the 1.3 cm continuum emission. The unit of the color bar is km s$^{-1}$.

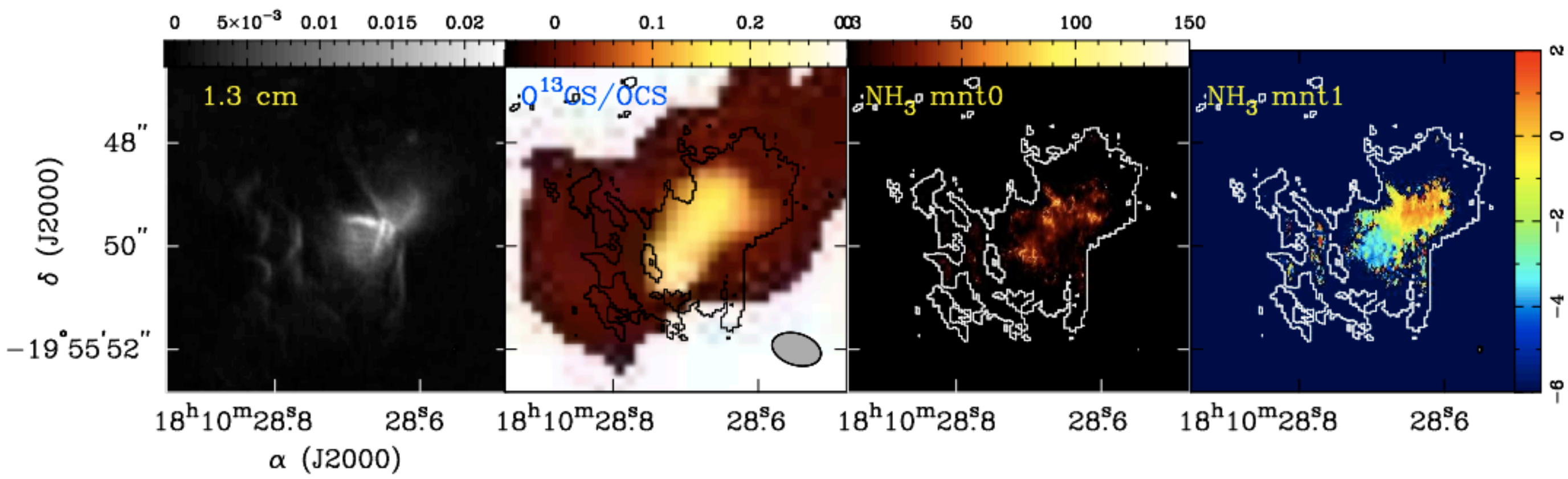

Figure 3: **Left:** VLA A-array 1.3 cm continuum map (Jy/beam). **Middle Left:** Ratio of the $O^{13}CS$ and OCS velocity integrated intensity. **Middle Right:** Velocity integrated $NH_3$ (3,3) satellite line optical depth ($\tau$·kms$^{-1}$). **Right:** Optical depth weighted averaged velocity of the $NH_3$ (3,3) satellite line (kms$^{-1}$). Contours show the 3% level of the 1.3 cm continuum emission. The VLA A-array 1.3 cm continuum and the $NH_3$ (3,3) satellite line data are reprocessed. See Sollins et al. (2005) and Sollins and Ho (2005) for the detailed configurations of these VLA observations.

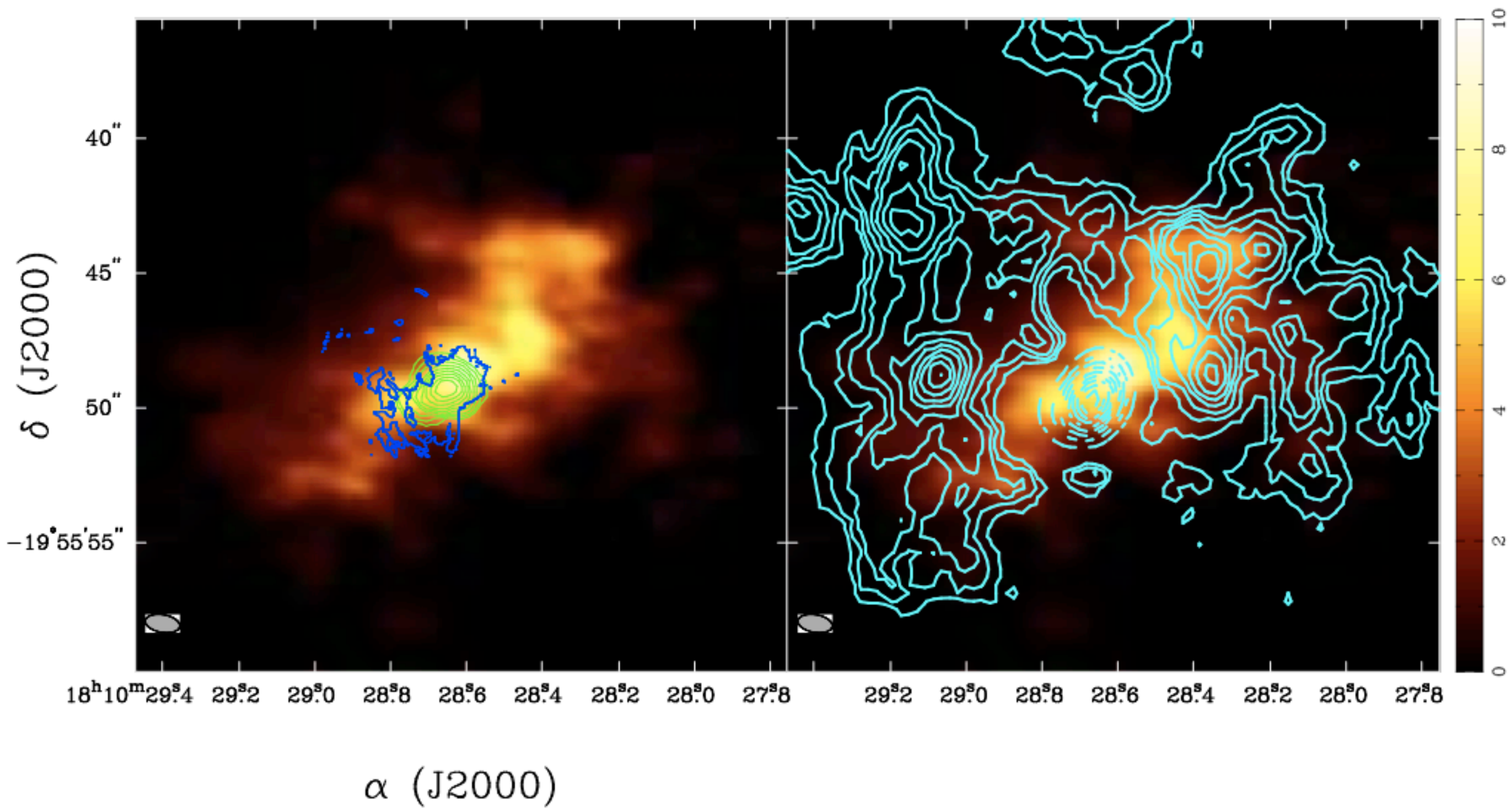

Figure 4: **Left:** Moment zero map of the $^{13}$CS J=5–4 line (color scale), the $CH_3CN$ J=12, K=5, 6 lines (green), and the 1.3 cm continuum image (blue). **Right:** Moment zero map of the $NH_3$ (3,3) main hyperfine emission line (cyan) and the $^{13}$CS J=5–4 line (color scale) The positive contours of the moment zero maps start from 10% of their emission peak value, with 10% steps (solid); the negative contours of the moment zero maps start from -200% of their emission peak value, with -200% steps (dashed). The 1.3 cm continuum is only shown in 3% peak value contour. The synthesized beam of the $^{13}$CS (5-4) map is shown at the bottom left of the image. The unit of the color bar is (Jy beam$^{-1}$)·(km s$^{-1}$).

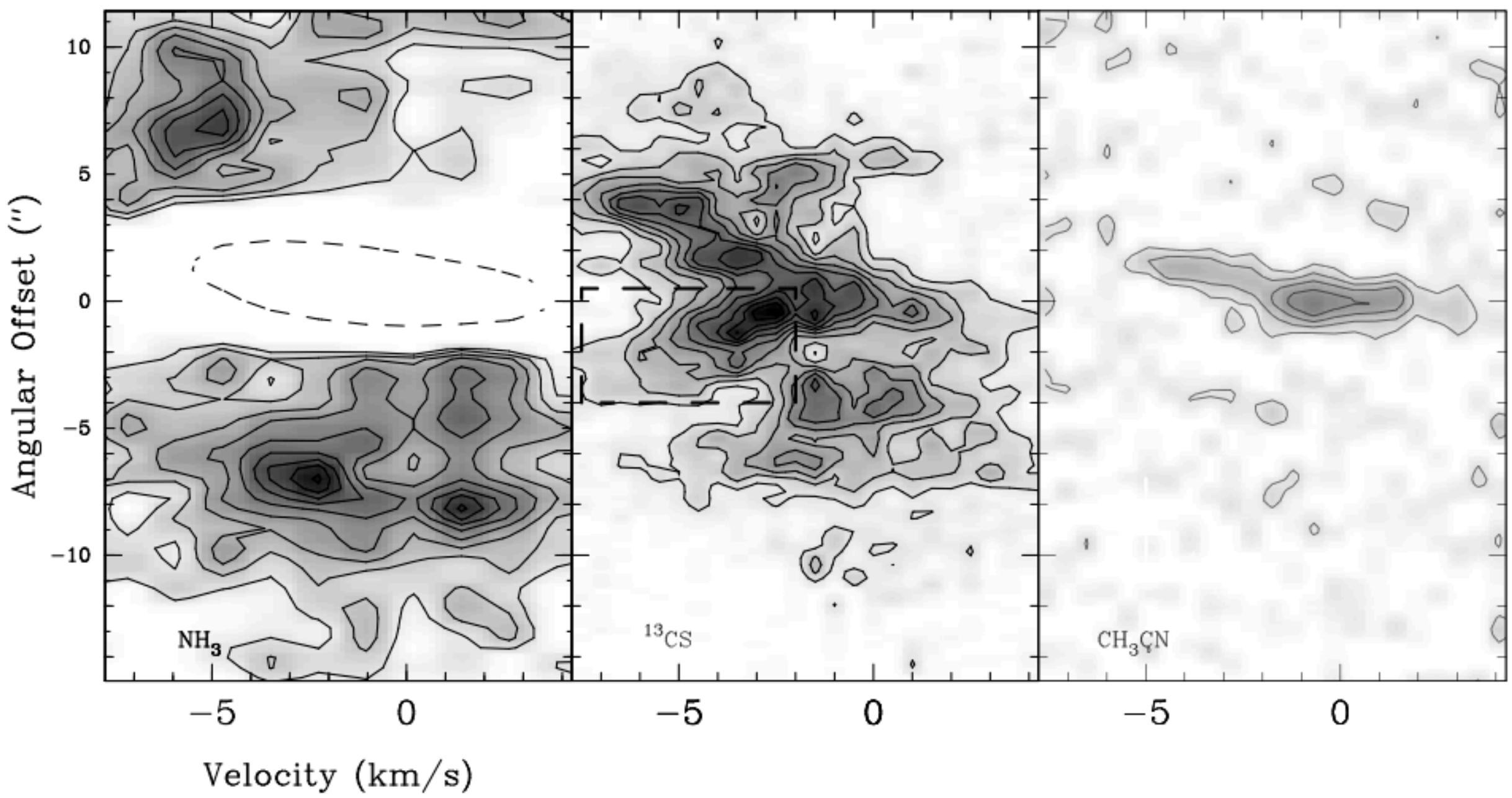

Figure 5: **Left:** The PV diagram of the $NH_3$ (3,3) main hyperfine line. Solid contours start from 10% of the emission peak, with 10% intervals. The dashed contour shows the level of -500% emission peak. Black line shows the velocity gradient of 14 $kms^{-1}pc^{-1}$. **Middle:** The PV diagram of the $^{13}$CS (5-4) line. Contours start from 10% of the emission peak, with 10% intervals. Black lines show the velocity gradient of 12 and 53 $kms^{-1}pc^{-1}$. The region enclosed by a dashed box in this panel is excluded while performing linear fittings of the velocity gradients. **Right:** The PV diagram of the $CH_3CN$ line. Contours start from 20% of the emission peak, with 20% intervals. Black lines show the velocity gradient of 100 and 133 $kms^{-1}pc^{-1}$. All PV cuts are centered at the coordinates of R.A. = $18^h10^m28^s.64$ and decl. = $-19^o55'49''.22$ with position angle P.A. = $140^o$ .

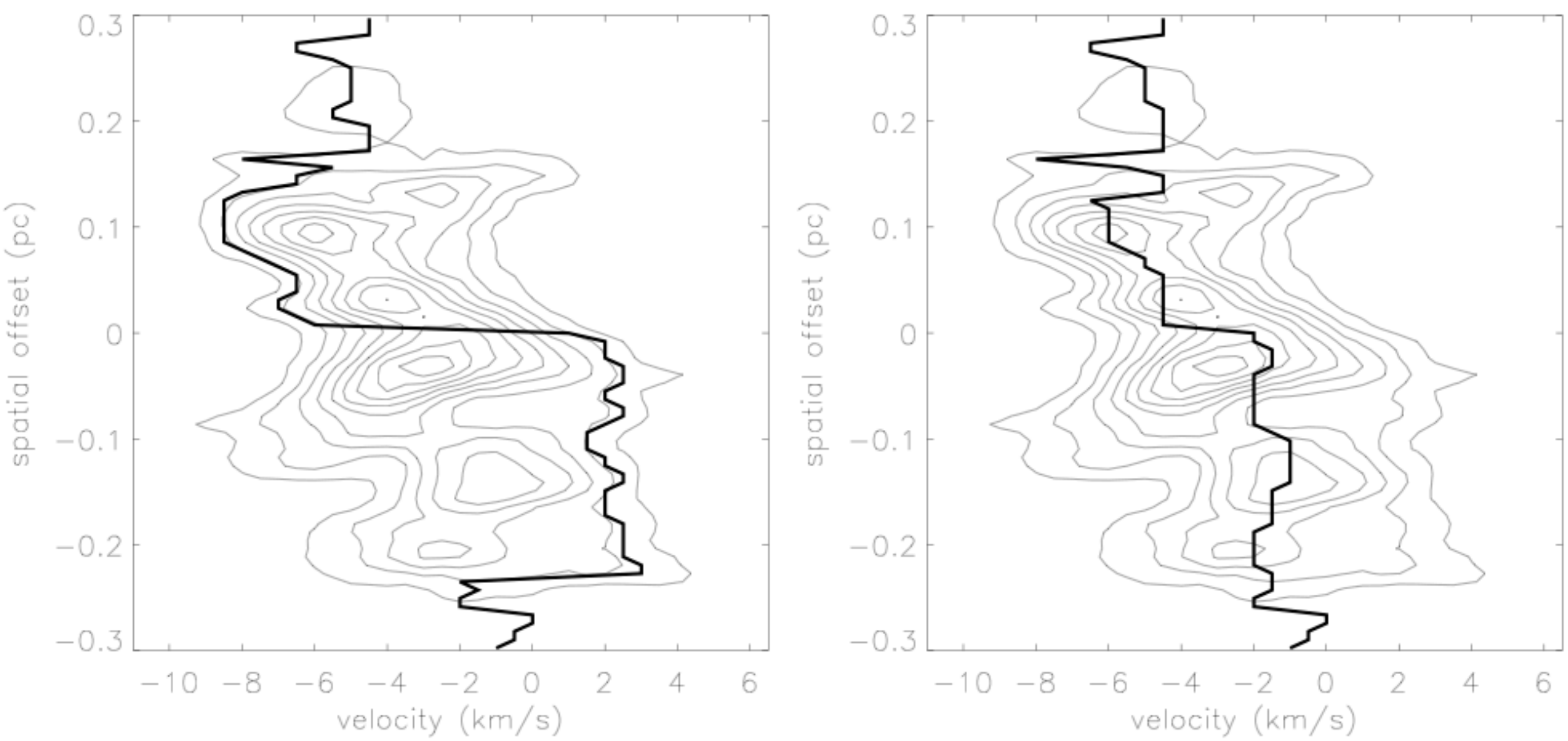

Figure 6: Examples of the results terminal velocity tracing method on $^{13}$CS (5-4) PV diagram. The terminal velocities defined by $\eta = 0.2$ (left) and $\eta = 0.8$ (right) are presented in black lines.

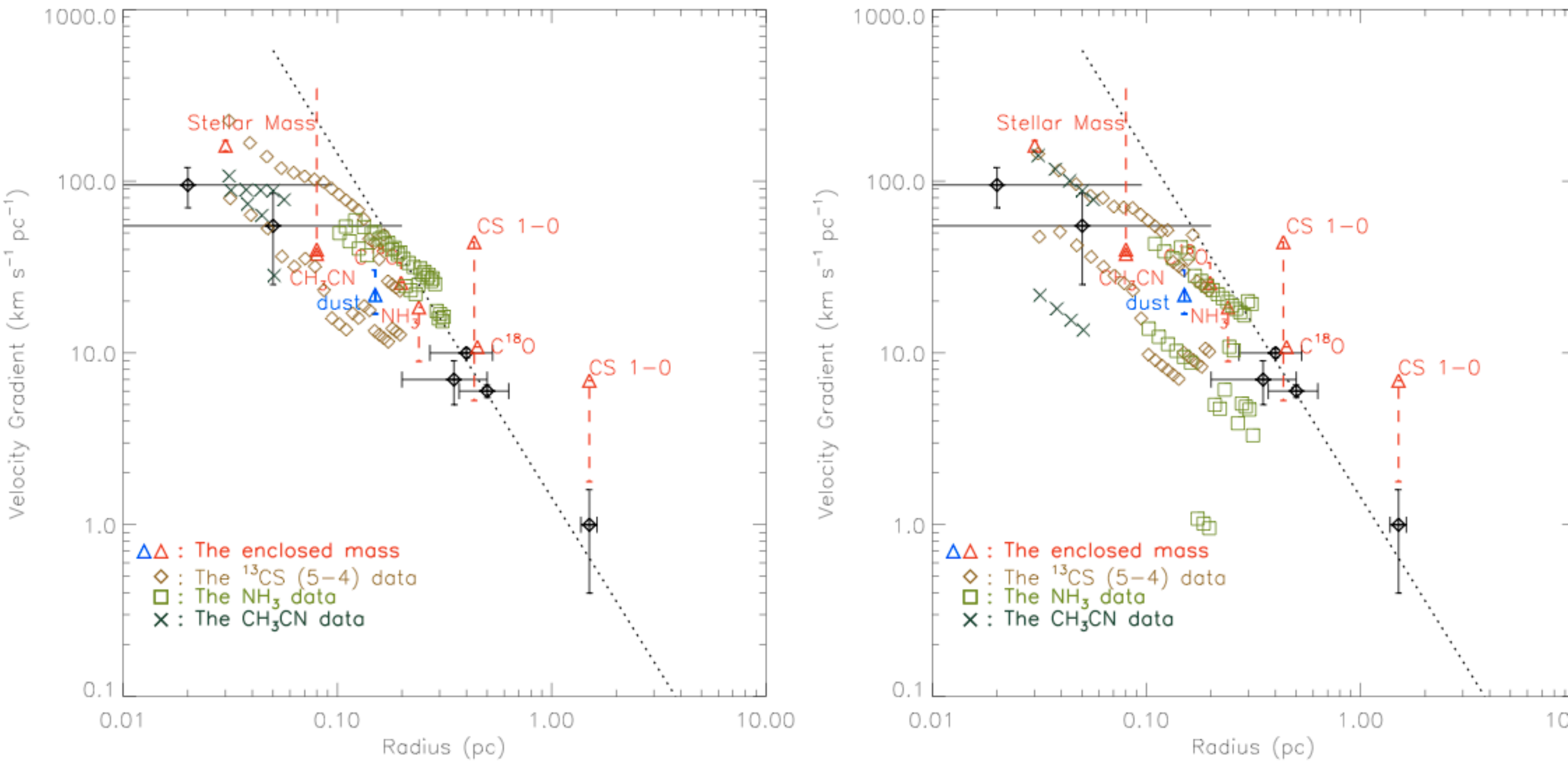

Figure 7: Velocity gradient at different radii. The dark green crosses represent the $CH_3CN$ data; the orange diamonds represent the $^{13}CS$ (5–4) data; and the green squares represent the $NH_3$ (3,3) main hyperfine inversion line data. The values in the left panel and the right panel are estimated using $\eta = 0.2$ and $\eta = 0.8$, respectively. **Black:** Velocity gradients at different radii reported by previous observations (From left to right: $NH_3$ (3,3) main hyperfine line (Keto, Ho & Haschick 1988); $NH_3$ (3,3) main hyperfine line (keto, Ho & Haschick 1987); $NH_3$ (3,3) main hyperfine line (Keto, Ho & Haschick 1987); $NH_3$ (1,1) main hyperfine line (Ho & Haschick 1986); CS (1-0) (Omodaka et al. 1992); $NH_3$ (1,1) main hyperfine line (Keto, Ho & Haschick 1987) ). **Red:** The necessary velocity gradients to balance the enclosed stellar and molecular mass estimated by the labeled molecules. Dashed error–bars take into account the uncertainties caused by abundance variations. We quote the estimated embedded stellar mass of 175 $M_\odot$ from Sollins et al. (2005). At the 1.5 pc radius, we provide the limit of the velocity gradient assuming that the embedded mass is 2500–40000 $M_\odot$. **Blue:** The necessary velocity gradients to balance the embedded stellar and molecular mass estimated by the 1.3 mm continuum. The dotted line sketches the expected velocity gradient assuming that the measured specific angular momentum at the radius of 0.5 pc is conserved.

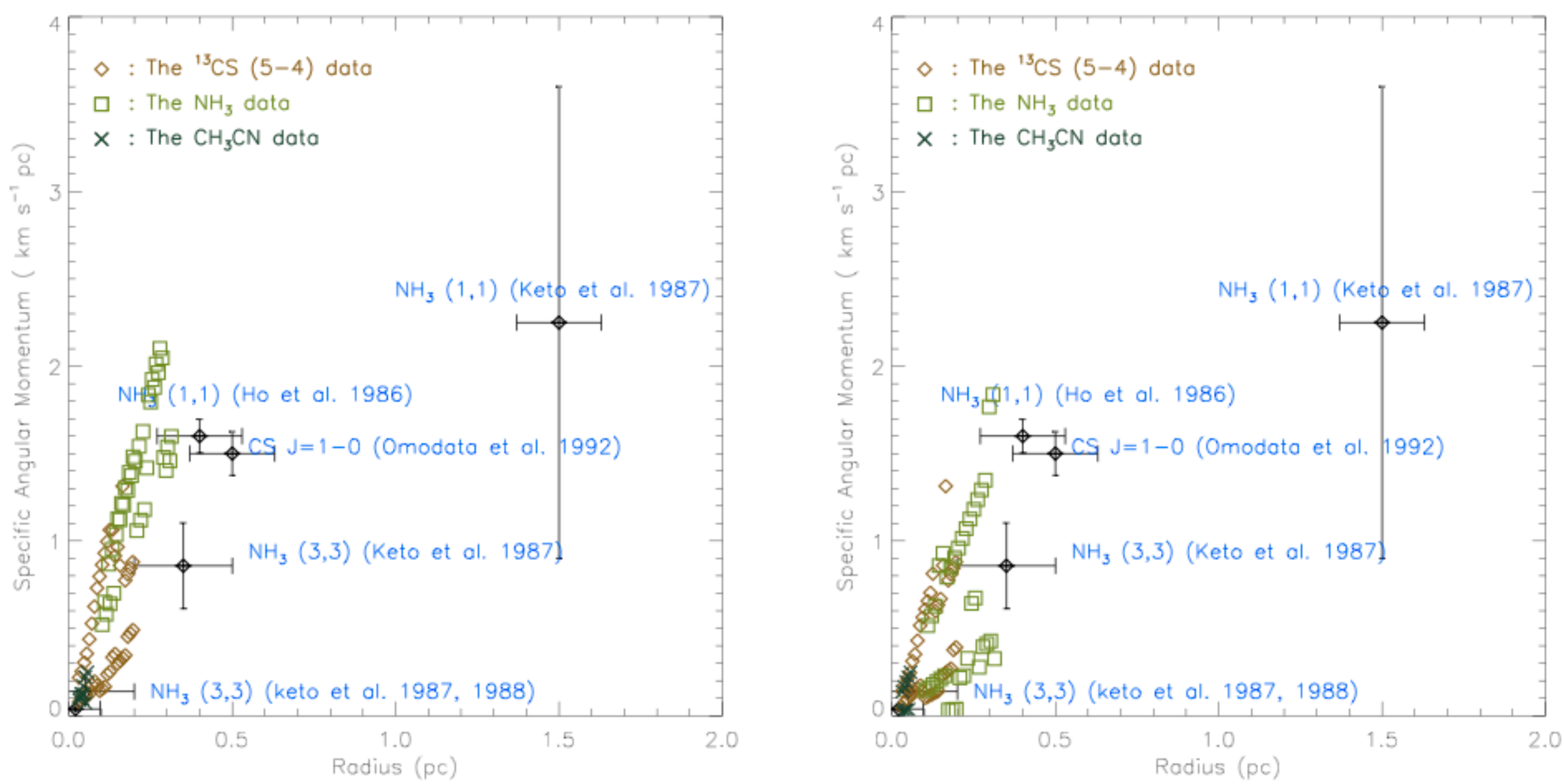

Figure 8: Specific angular momentum at different radii traced by the available molecular line data. The dark green cross represent the $CH_3CN$ data; the orange diamonds represents the $^{13}CS$ (5–4) data; and the green squares represent the $NH_3$ (3,3) main hyperfine inversion line data. The values in the left panel and the right panel are estimated using $\eta = 0.2$ and $\eta = 0.8$, respectively. Black circles show the values reported by previous observations.